\documentclass[12pt]{article}

\usepackage{graphics}
\usepackage{epsfig}
\usepackage{cite}
\input epsf

\title{Open $b$ production at LHC and Parton Shower Effects}

\author{H.~Jung$^{1,2}$, M.~Kraemer$^1$, A.V.~Lipatov$^3$, N.P.~Zotov$^3$}

\begin{document}
\maketitle

\vspace*{-7.5cm}
\begin{flushright}
DESY 11-086\\
May 2011 \\
\end{flushright}
\vspace*{+4.5cm}

\begin{center}

{\it $^1$DESY, Hamburg, Germany\\[3mm]
$^2$ CERN, Geneva, Switzerland\\
University of Antwerp, Antwerp, Belgium\\[3mm]
$^3$D.V.~Skobeltsyn Institute of Nuclear Physics,\\ 
M.V. Lomonosov Moscow State University, Russia}\\[3mm]

\end{center}

\vspace{0.5cm}

\begin{center}

{\bf Abstract }

\end{center}

We present hadron-level predictions from the Monte Carlo generator {\sc Cascade} and 
numerical level calculations of beauty quark and inclusive $b$-jet production in the framework of 
the $k_T$-factorization QCD approach for CERN LHC energies. 
The unintegrated gluon densities in a proton are determined using the CCFM 
evolution equation and the Kimber-Martin-Ryskin (KMR) prescription.
We study the theoretical uncertainties of our calculations and investigate the 
effects coming from parton showers in initial and final states.
Our predictions are compared with the recent  data taken by 
the CMS collaboration.

\vspace{0.8cm}

\noindent
PACS number(s): 12.38.-t, 13.85.-t

\vspace{0.5cm}

\section{Introduction} \indent 

Beauty production at high energies is subject of intense 
studies from both theoretical and experimental 
points of view since events containing $b$ quarks present an important 
background to many of the searches at the LHC.
From the theoretical point, the dominant production mechanism is believed 
to be quark pair production through the gluon-gluon fusion subprocess and
therefore these processes provide an opportunity to test the different 
predictions based on Quantum Chromodynamics (QCD).
The present note is motivated by the recent measurements \cite{1, 2}
 $b$ production performed by the 
CMS collaboration. The $b$-quark cross sections 
have been presented~\cite{1} as a function of the muon transverse momentum and pseudorapidity 
at $\sqrt s = 7$~TeV. It was observed that the data tend to be 
higher than the \textsc{MC@NLO} \cite{3, 4} predictions.
On the other hand the measurements of inclusive
$b$-jet production cross sections \cite{2} are reasonably well described 
by \textsc{MC@NLO}. In addition to the comparison of CASCADE with data 
in \cite{1} we present here futher studies.

In the framework of the $k_T$-factorization approach of QCD \cite{5}, 
which is of primary consideration in this note, a study of the heavy quark 
production has been done (for previous results see \cite{6, 7, 8, 9, 
10, 11, 12}).
In our previous study \cite{12} we show a good agreement between the Tevatron data on the $b$ quarks, $b \bar b$ di-jets, 
$B^+$ and several $D$ mesons (or rather muons from their semileptonic decays) production with the predictions coming from $k_T$-factorization and we investigated  the role of initial and final state parton showers.
Based on these results, we give here
a systematic analysis of the recent CMS measurements~\cite{1,2} in the framework of  $k_T$-factorization. 
As done in \cite{12}, we produce the calculations in two ways:
we will perform numerical parton-level calculations  (labeled as LZ) as well as calculations with the full hadron level Monte Carlo event generator 
\textsc{Cascade}~\cite{13} and compare both with the measured cross sections of heavy quark production. 
In this way we will investigate the influence of parton showers in 
initial and final states for the description of the data. 
Additionally we study different sources of theoretical uncertainties, i.e. uncertainties connected with the gluon evolution scheme, 
heavy quark mass, hard scale 
of partonic subprocess and the heavy quark fragmentation functions. 

The outline of our paper is the following. In Section~2 we 
recall very shortly the basic formulas of the $k_T$-factorization approach with a brief 
review of calculation steps. In Section~3 we present the numerical results
of our calculations and a discussion. Section~4 contains our conclusions.

\section{Theoretical framework} \indent 

In the present analysis we follow the approach
described in the earlier publication \cite{12}.
For the reader's convenience, we only briefly recall
here main points of the theoretical scheme.

The cross section of heavy quark hadroproduction 
at high energies in the $k_T$-factorization approach
is calculated as a convolution of the off-shell (i.e. $k_T$-dependent)
partonic cross section $\hat \sigma$ and the unintegrated gluon 
distributions in a proton. It can be presented in the following form:
$$
  \displaystyle \sigma (p \bar p \to Q\bar Q \, X) = \int {1\over 16\pi (x_1 x_2 s)^2 } {\cal A}(x_1,{\mathbf k}_{1T}^2,\mu^2) {\cal A}(x_2,{\mathbf k}_{2T}^2,\mu^2) |\bar {\cal M}(g^* g^* \to Q\bar Q)|^2 \times \atop
  \displaystyle  \times d{\mathbf p}_{1T}^2 d{\mathbf k}_{1T}^2 d{\mathbf k}_{2T}^2 dy_1 dy_2 {d\phi_1 \over 2\pi} {d\phi_2 \over 2\pi}, \eqno (1)
$$

\noindent 
where ${\cal A}(x,{\mathbf k}_{T}^2,\mu^2)$ is the
unintegrated gluon distribution in a proton, 
$|\bar {\cal M}(g^* g^* \to Q\bar Q)|^2$ is the 
off-shell (i.e. depending on the initial gluon virtualities 
${\mathbf k}_{1T}^2$ and ${\mathbf k}_{2T}^2$) matrix element squared 
and averaged over initial gluon 
polarizations and colors, and 
$s$ is the total center-of-mass energy.
The produced heavy quark $Q$ and anti-quark $\bar Q$ have the 
transverse momenta ${\mathbf p}_{1T}$ and
${\mathbf p}_{2T}$ and the center-of-mass rapidities $y_1$ and $y_2$.
The initial off-shell gluons have a fraction $x_1$ and $x_2$ 
of the parent protons longitudinal 
momenta, non-zero transverse momenta ${\mathbf k}_{1T}$ and 
${\mathbf k}_{2T}$ (${\mathbf k}_{1T}^2 = - k_{1T}^2 \neq 0$, 
${\mathbf k}_{2T}^2 = - k_{2T}^2 \neq 0$) and azimuthal angles
 $\phi_1$ and $\phi_2$. 
The analytic expression for the 
$|\bar {\cal M}(g^* g^* \to Q\bar Q)|^2$ can be found, for example, in~\cite{5, 9}.

The unintegrated gluon distribution in a 
proton ${\cal A}(x,{\mathbf k}_{T}^2,\mu^2)$ in (1)
can be obtained from the analytical or numerical solution of the 
Balitsky-Fadin-Kuraev-Lipatov (BFKL)~\cite{14} or 
Ciafaloni-Catani-Fiorani-Marchesini (CCFM)~\cite{15} equations.
As in \cite{12}, in the numerical calculations we have tested 
a few different sets, namely CCFM A0 (B0)~\cite{16} and KMR~\cite{17} ones. 
The input parameters in both CCFM-evolved gluon densites 
have been fitted \cite{16} to describe the proton structure function $F_2(x, Q^2)$.
The difference between A0 and B0 sets is connected with the different values of 
soft cut and width of the intrinsic ${\mathbf k}_{T}$ distribution.
A reasonable description of the $F_2$ data
can be achieved~\cite{16} by both these sets.
To evaluate the unintegrated gluon densities in a proton ${\cal A}(x,{\mathbf k}_T^2,\mu^2)$ 
we apply also the Kimber-Martin-Ryskin (KMR) approach~\cite{17}. The KMR approach is a 
formalism to construct the unintegrated parton (quark and gluon) distributions from the 
known conventional parton distributions. For the input, we have used the 
standard GRV 94~(LO)~\cite{18}
 (in LZ calculations) and 
MRST 99~\cite{20} (in \textsc{Cascade}) sets.

\section{Numerical results} \indent

The unintegrated gluon distributions to be used in the cross section (1) depend on
the renormalization and factorization scales $\mu_R$ and $\mu_F$.
Following \cite{12}, in the numerical calculations we set 
$\mu_R^2 = m_Q^2 + ({\mathbf p}_{1T}^2 + {\mathbf p}_{2T}^2)/2$,
$\mu_F^2 = \hat s + {\mathbf Q}_T^2$, where ${\mathbf Q}_T$ is the 
transverse momentum of the initial off-shell gluon pair,
$m_c = 1.4 \pm 0.1$~GeV, $m_b = 4.75 \pm 0.25$~GeV.  We use the LO formula 
for the coupling $\alpha_s(\mu_R^2)$ with $n_f = 4$ active quark flavors
at $\Lambda_{\rm QCD} = 200$~MeV, such 
that $\alpha_s(M_Z^2) = 0.1232$. 

We begin the discussion by presenting our results for the 
muons originating from the semileptonic decays of the $b$ quarks. 
The CMS collaboration has measured~\cite{1} the transverse momentum and 
pseudorapidity distributions of  muons from $b$-decays. The measurements have been performed 
in the kinematic range
$p_T^\mu > 6$~GeV and $|\eta^\mu| < 2.1$ at the total center-of-mass energy $\sqrt s = 7$~TeV.
To produce muons from $b$-quarks in the LZ calculations, we first convert $b$-quarks into $B$ mesons 
using the Peterson fragmentation function with default value $\epsilon_b = 0.006$ 
and then simulate their semileptonic decay according to the standard electroweak theory.
The branching of $b \to \mu $ as well as the cascade decay $b\to c\to \mu$ are taken into 
account with the relevant branching fractions taken from~\cite{22}.
The predictions of the LZ and {\sc Cascade} calculations are shown in 
Figs.~1 and~2 in comparison with the CMS data.
We find a good description of the data when using 
the CCFM-evolved (A0) gluon distribution in LZ calculations
although the {\sc Cascade} curves tend to lie slightly below the data at central 
rapidities. 
The predictions between the LZ  and \textsc{Cascade} calculations agree well at parton level.
The observed difference between them in Figs.~1 and~2 is due 
to missing parton shower effects in the LZ calculations.
The influence of such effects is demonstrated in Fig.~3, where we 
show separately the results of our \textsc{Cascade} calculations without parton 
shower, with only initial state, with only final state and with 
both initial and final state parton showers. 
One can see that without
initial and final state parton showers, the \textsc{Cascade} predictions
are very close to the LZ ones.
The similar situation was pointed out in~\cite{12} at the Tevatron energies.

To investigate the dependence of our predictions on the quark-to-hadron fragmentation function,
we repeated our calculations with the shifted value of the Peterson 
shape parameter $\epsilon_b = 0.003$, which is 
is often used in the NLO pQCD calculations.
Additionally, we have applied the non-perturbative fragmentation functions which
have been proposed in~\cite{23, 24, 25}. The input parameters 
were determined~\cite{24, 25} by a  fit to LEP data. The results of our calculations
are shown in Fig.~4. For illustration, we used here the CCFM A0 gluon density.
We find that the predicted cross sections 
in the considered kinematic region are larger for smaller values of the 
parameter $\epsilon_b$ or if the fragmentation function from~\cite{23, 24, 25} is used.
Thus, the CMS data points lie within the band of theoretical uncertainties. 
The results obtained here (see Fig.~1) with the CCFM B0 and KMR gluon densities (but 
also with A0 density as shown in the CMS paper) 
 are rather close to the {\textsc MC@NLO} ones (not shown) 
and underestimate the data by a factor of 1.6.

\begin{table}
\begin{center}
\begin{tabular}{|l|c|}
\hline
   & \\
  Source & $\sigma(pp \to b + X \to \mu + X^\prime,$ $p_T^\mu > 6$~GeV, $|\eta^\mu| < 2.1)$ \\
   & \\
\hline
   & \\
   CMS data [$\mu$b]  & $1.32 \pm 0.01$ (stat) $\pm 0.30$ (syst) $\pm 0.15$ (lumi)\\
   & \\
   \hline
   & \\
   A0 (LZ/\textsc{Cascade}) & 1.31/0.96 \\
   & \\
   B0 (LZ/\textsc{Cascade}) & 0.98/0.72 \\
   & \\
   KMR (LZ/\textsc{Cascade}) & 0.91/0.59 \\
   & \\
   \textsc{MC@NLO} \protect\cite{1} & 0.95 \\
   & \\
   \textsc{Pythia}  \protect\cite{1} & 1.9 \\
   & \\
   \hline
\end{tabular}
\end{center}
\caption{The inclusive b-quark production cross section in 
  $pp$ collisions at $\sqrt s = 7$~TeV.}
\label{table_pdfs}
\end{table}

\begin{table}
\begin{center}
\begin{tabular}{|l|c|}
\hline
   & \\
  Source & $\sigma(pp \to b + X \to \mu + X^\prime,$ $p_T^\mu > 6$~GeV, $|\eta^\mu| < 2.1)$ \\
   & \\
\hline
   & \\
   CCFM set A0 & 0.96 $\mu $b\\
   & \\
   CCFM set A0$+$ & +13\% \\
   & \\
   CCFM set A0$-$ & -2\% \\
   & \\
   $m_b=5.0$ GeV & -7\% \\
   & \\
   $m_b=4.5$ GeV & +6\% \\
   & \\
   $\epsilon_b=0.003$ & +9\% \\
   & \\
   \hline
   & \\
   Total & $\pm^{17\%}_{7\%}$ \\
   & \\
\hline
\end{tabular}
\end{center}
\caption{Systematic uncertainties for beauty total cross section in 
  $p p$ collisions at $\sqrt s = 7$~TeV obtained with \textsc{Cascade}.}
\label{table_uncertainties}
\end{table}

The visible cross sections of decay muons from $b$-decays
are listed in Table~1 in comparison with the CMS data~\cite{1}. In Table~2
the systematic uncertainties of our calculations are summarized.
To estimate the uncertainty coming from the renormalization scale $\mu_R$, we used 
the CCFM set A0$+$ and A0$-$ instead of the default density function A0.
These two sets represent a variation of the scale used in $\alpha_s$ in the 
off-shell matrix element. The A0$+$ stands for a variation of $2\,\mu_R$, 
while set A0$-$ reflects $\mu_R/2$. 
We observe a deviation of roughly $13\%$ for set A0$+$.
The uncertainty coming from set A0$-$ is generally smaller and negative.
The dependence on the $b$-quark mass is investigated by variation of $b$-quark mass of  $m_b=4.75$~GeV by $\pm0.25$ GeV. 
The calculated $b$-quark cross sections vary by $\sim \pm 6 \%$.

The CMS collaboration has measured~\cite{2} the double differential cross
sections $d\sigma/dy\,dp_T$ of inclusive $b$-jet production at the $\sqrt s = 7$~TeV.
The measurements have been determined in four $b$-jet rapidity regions,
namely $|y| < 0.5$, $0.5 < |y| < 1$, $1 < |y| < 1.5$ and $1.5 < |y| < 2$. 
Our predictions are shown in Figs.~5 and 6 and compared to the CMS data.
In the \textsc{Cascade} calculations the $b$-jets are reconstructed with the 
anti-$k_t$ cone algorithm \cite{Cacciari:2008gp} (using the {\sc Fastjet} package \cite{Cacciari:2005hq,fastjet}) with radius $R > 0.5$. 
In contrast with the decay muon cross sections, 
the predictions based on the CCFM and KMR gluons are very similar to each other.
The reasonable description of the data is obtained by all unintegrated 
gluon densities under consideration.

Finally, we would like to point out the 
role of non-zero gluon transverse 
momentum $k_T$ in the off-shell matrix elements (see Figs.~7 and 8).
In these Figs, the solid histograms correspond to the results obtained according to the 
master formula~(1). The dotted histograms are obtained by using the same formula
but without virtualities of the incoming gluons in partonic amplitude
and with the additional requirement ${\mathbf k}_{1,2 \,T}^2 < \mu_R^2$.
We find that the non-zero gluon transverse momentum
in the hard matrix element is important for the description of data.
The similar situation was pointed out in~\cite{12} at the Tevatron energies.
It means, that the high ${\mathbf k}_{T}$ region is important, and only when 
including the high${\mathbf k}_{T}$ tail the results are similar to NLO predictions. 

\section{Conclusions} \indent 

In this note we analyzed the first data on the beauty production in 
$pp$ collisions at LHC taken by the CMS collaboration.
Our consideration is based on the $k_T$-factorization 
approach supplemented with the CCFM-evolved unintegrated gluon
densities in a proton. The analysis covers the total and differential cross sections 
of muons originating from the semileptonic decays of  beauty quarks as well as
the double differential cross sections of inclusive $b$-jet production. 
Using the full hadron-level Monte Carlo generator {\sc Cascade}, we investigated the 
effects coming from the parton showers in initial and final states.
Different sources of theoretical uncertainties have been studied.

Our LZ predictions with the default set of parameters agree with the data.
The \textsc{Cascade} predictions tend to slightly 
underestimate the data at central rapidities but the data points still lie within the band of 
theoretical uncertainties. In this case the overall description of the data at a 
similar level of agreement as in the framework of NLO collinear QCD factorization.

\section{Acknowledgments} \indent 
We are grateful to comments from V.~Chiochia, G.~Dissertori, W.~Erdmann, V.~Zhukov and
S.~Baranov. The authors are very grateful to 
DESY Directorate for the support in the 
framework of Moscow --- DESY project on Monte-Carlo
implementation for HERA --- LHC.
A.V.L. was supported in part by the Helmholtz --- Russia
Joint Research Group and by the grant of president of 
Russian Federation (MK-3977.2011.2).
Also this research was supported by the 
FASI of Russian Federation (grant NS-4142.2010.2),
FASI state contract 02.740.11.0244 and 
RFBR grant 11-02-01454-a.

\newpage

\begin{figure}
\epsfig{figure=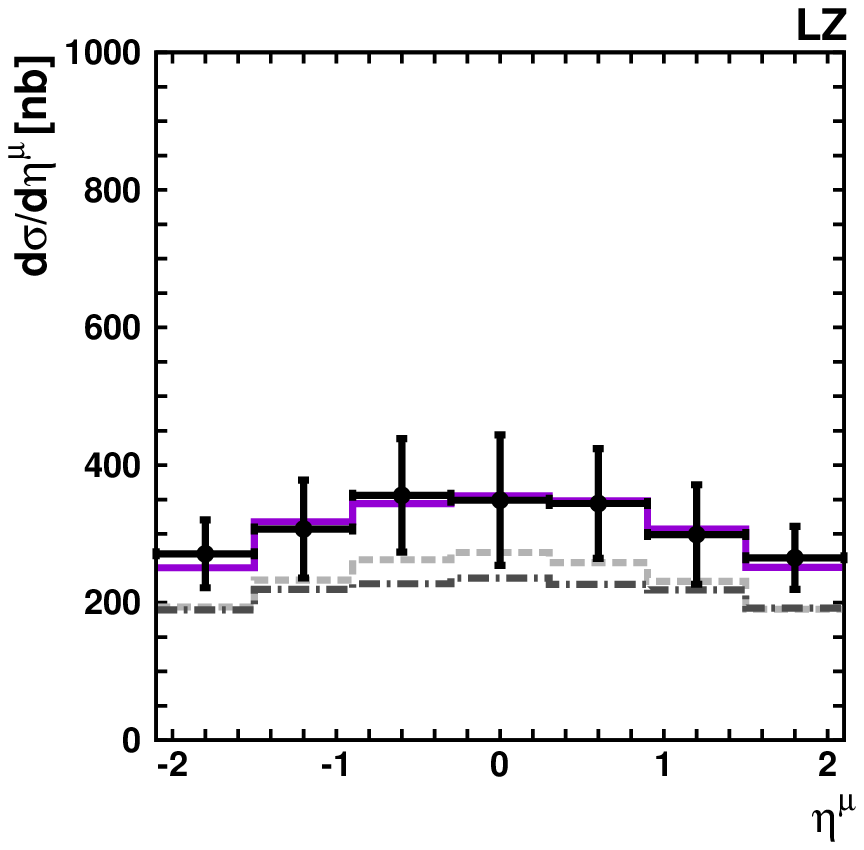, width = 8.1cm}
\put(0.0,-0.4){\epsfig{figure=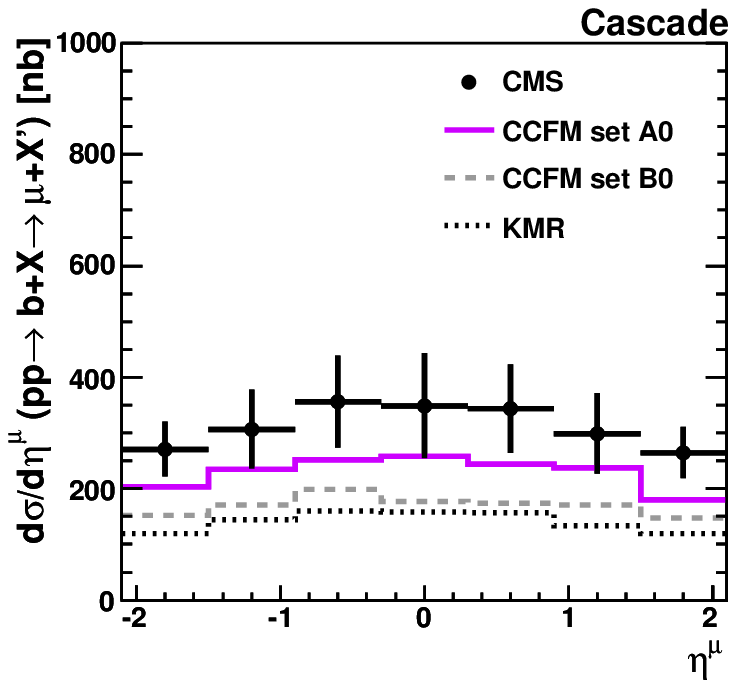, width = 7.5cm}}
\caption{The pseudorapidity distributions of muons arising from the 
semileptonic decays of beauty quarks. The first column shows the LZ numerical 
results while the second one depicts the \textsc{Cascade} predictions. 
The solid, dashed and dotted histograms 
correspond to the results obtained with the CCFM A0, B0
and KMR unintegrated gluon densities.
The kinematic cuts applied are described in the text.
The experimental data are from CMS \cite{1}.}
\label{fig1}
\end{figure}

\begin{figure}
\epsfig{figure=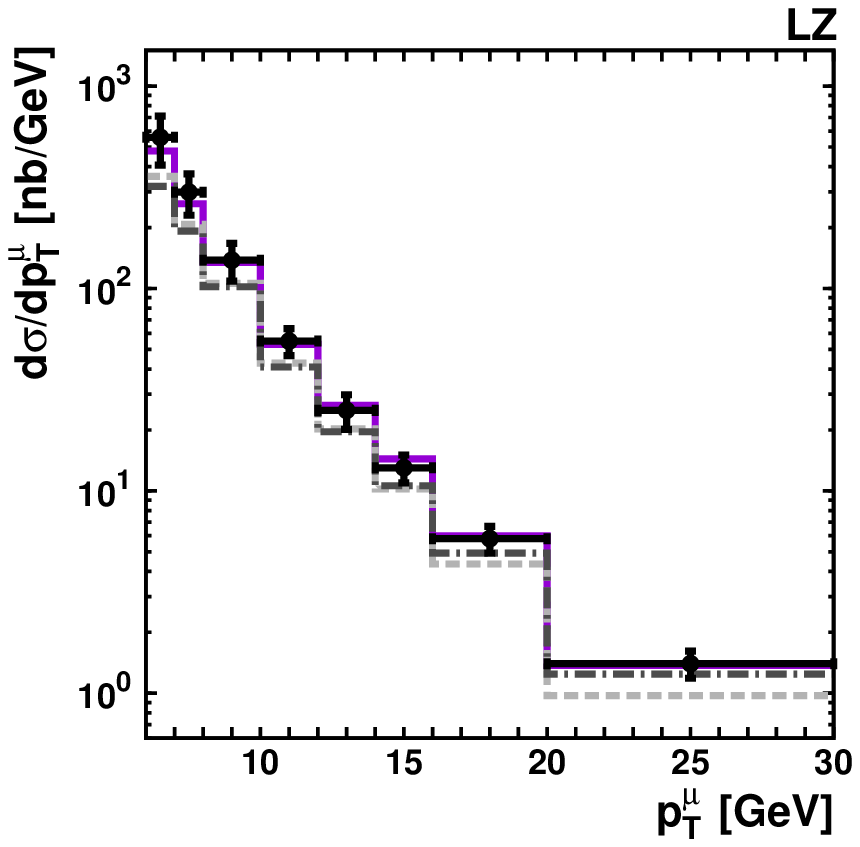, width = 8.1cm}
\put(0.0,-0.4){\epsfig{figure=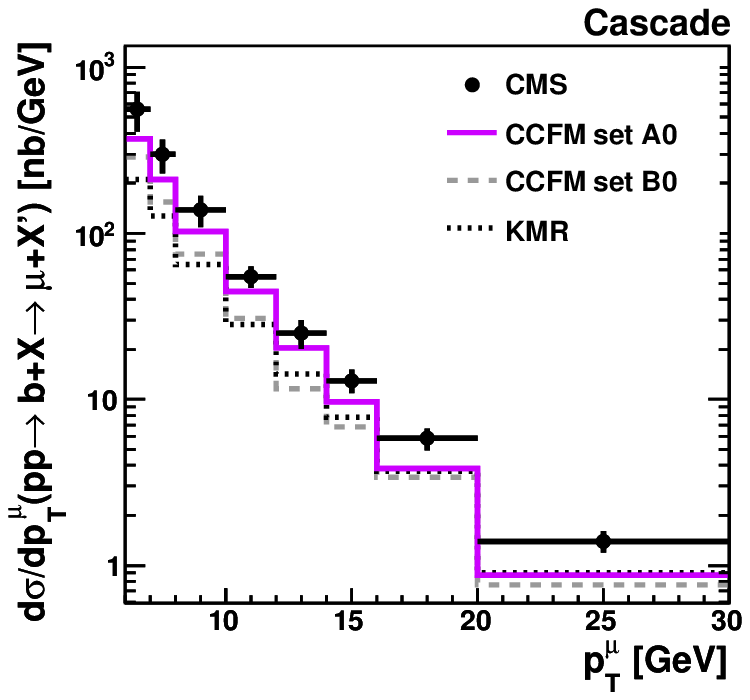, width = 7.5cm}}
\caption{The transverse momentum distributions of muons arising from the 
semileptonic decays of beauty quarks. The first column shows the LZ numerical 
results while the second one depicts the \textsc{Cascade} predictions. 
Notation of all histograms is the same as in Fig.~1.
The kinematic cuts applied are described in the text.
The experimental data are from CMS \cite{1}.}
\label{fig2}
\end{figure}

\begin{figure}
\hspace{0.6cm}
\epsfig{figure=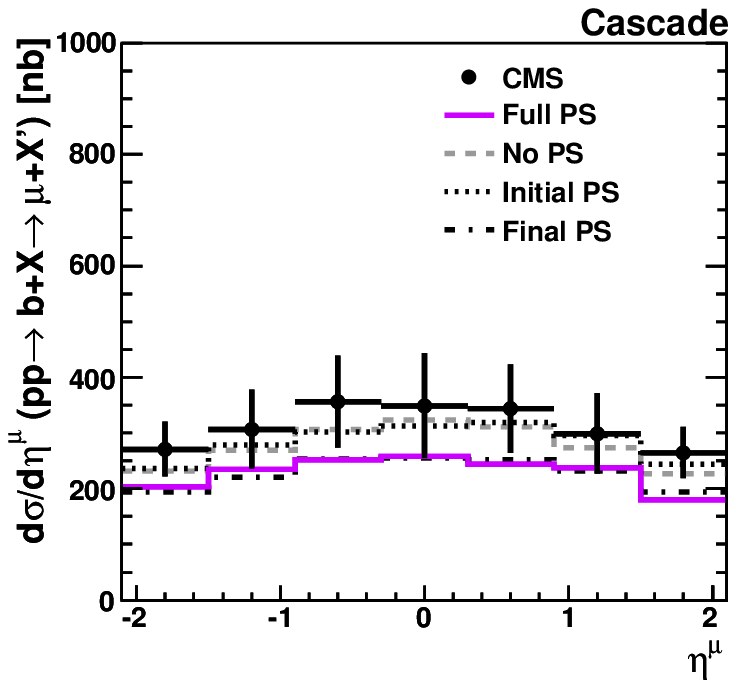, width = 7.5cm}
\hspace{0.2cm}
\epsfig{figure=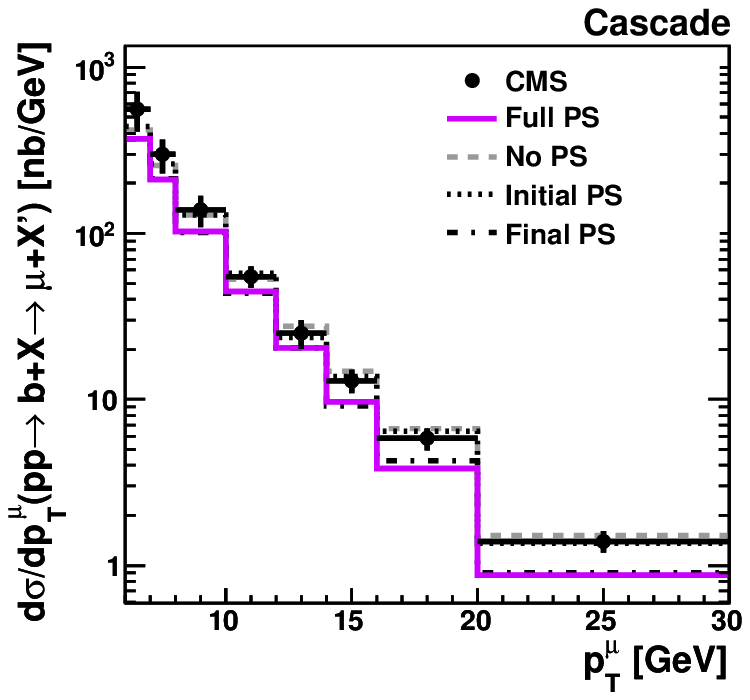, width = 7.5cm}
\caption{Parton shower effects in the pseudo-rapidity and transverse momentum 
distributions of the muons. The four lines represent full 
parton shower (solid line), no parton shower (dashed line), initial state
parton shower (dashed dotted line) and final state parton shower (dotted line).
The experimental data are from CMS \cite{1}.}
\label{fig3}
\end{figure}

\begin{figure}
\epsfig{figure=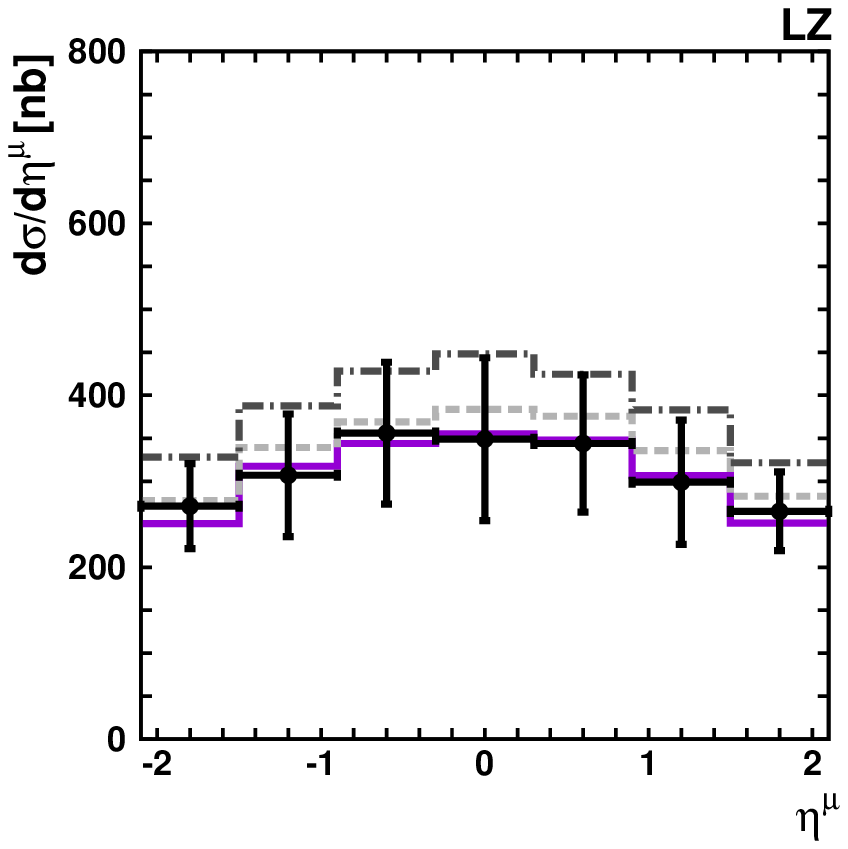, width = 8.1cm}
\epsfig{figure=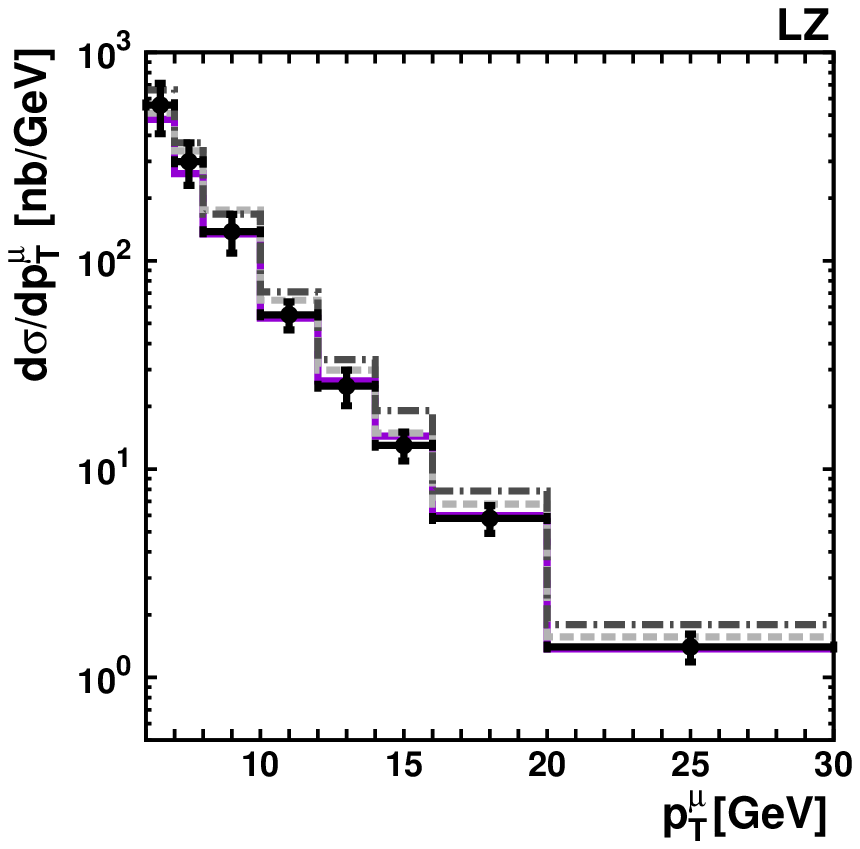, width = 8.1cm}
\caption{The dependence of our predictions on the fragmentation scheme.
The solid, dashed and dash-dotted histograms correspond to the results
obtained using the Peterson fragmentation function with $\epsilon_b = 0.006$, 
$\epsilon_b = 0.003$ and the non-perturbative fragmentation functions 
from~\cite{23, 24, 25}, respectively. We use CCFM-evolved (A0) gluon density for illustration.
The experimental data are from CMS \cite{1}.}
\label{fig4}
\end{figure}

\begin{figure}
\begin{center}
\epsfig{figure=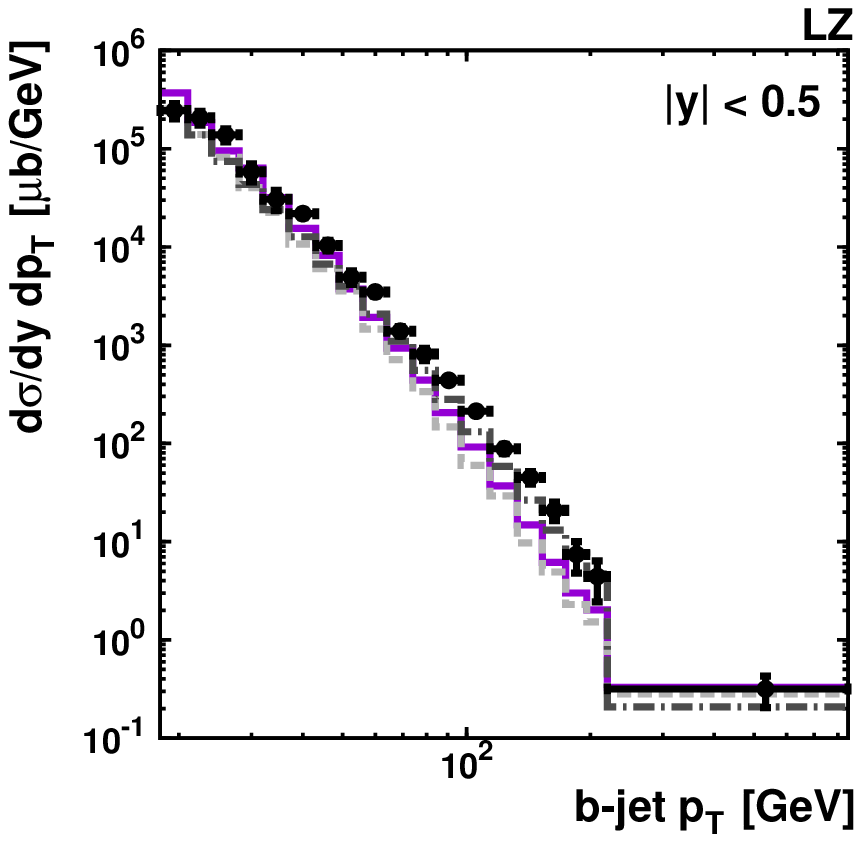, width = 8.1cm}
\epsfig{figure=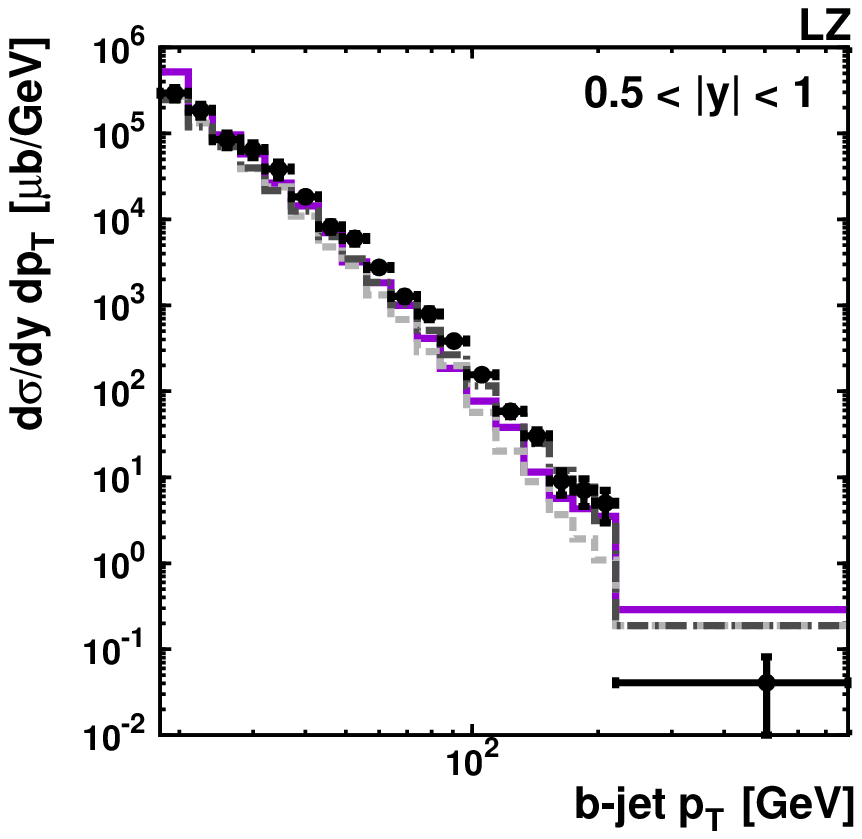, width = 8.1cm}
\epsfig{figure=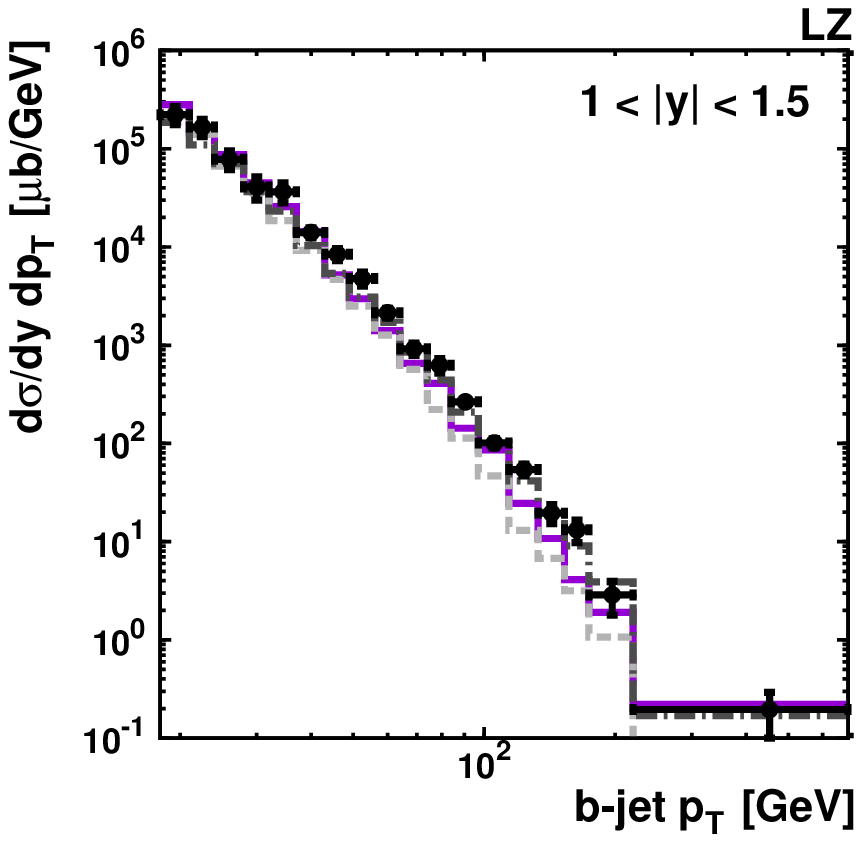, width = 8.1cm}
\epsfig{figure=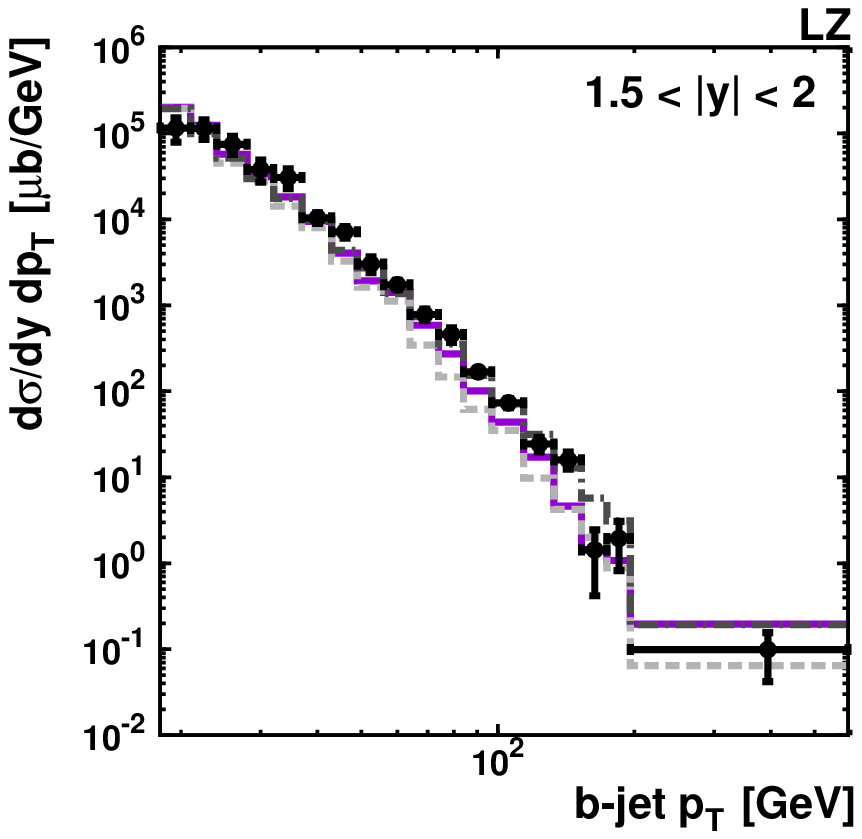, width = 8.1cm}
\caption{The double differential cross sections $d\sigma/dy\,dp_T$ of inclusive
$b$-jet production as a function of $p_T$ in different $y$ regions
calculated at $\sqrt s = 7$~TeV (LZ predictions). 
Notation of all histograms is the same as in Fig.~1.
The experimental data are from CMS~\cite{2}.}
\end{center}
\label{fig5}
\end{figure}

\begin{figure}
\begin{center}
\epsfig{figure=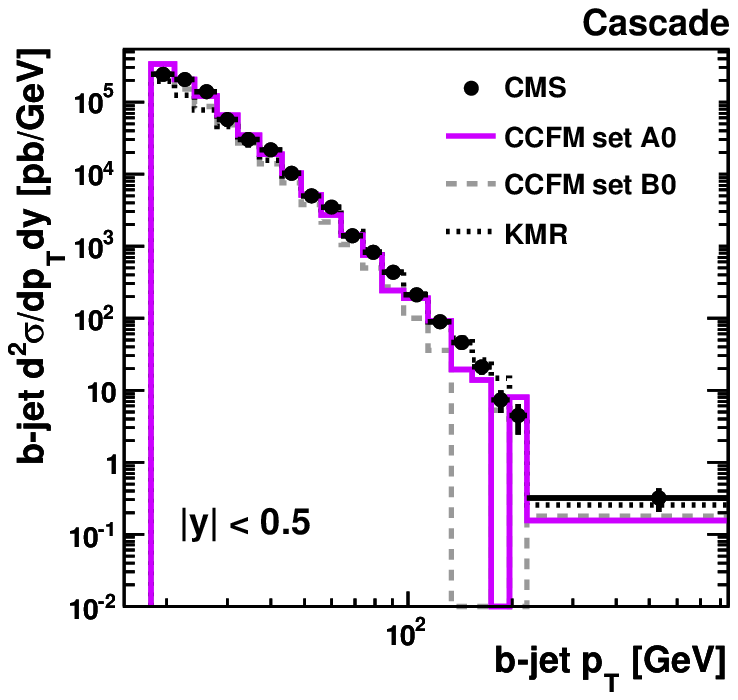, width = 8.1cm}
\epsfig{figure=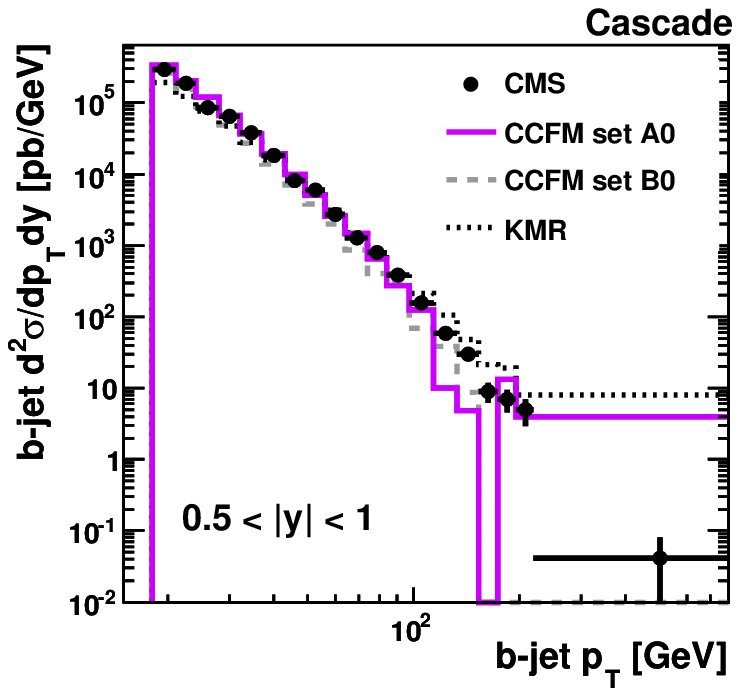, width = 8.1cm}
\epsfig{figure=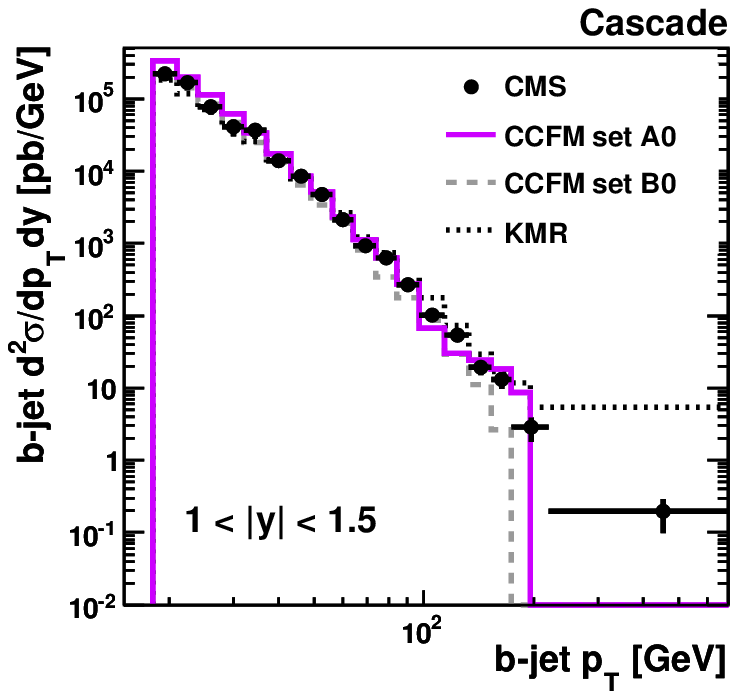, width = 8.1cm}
\epsfig{figure=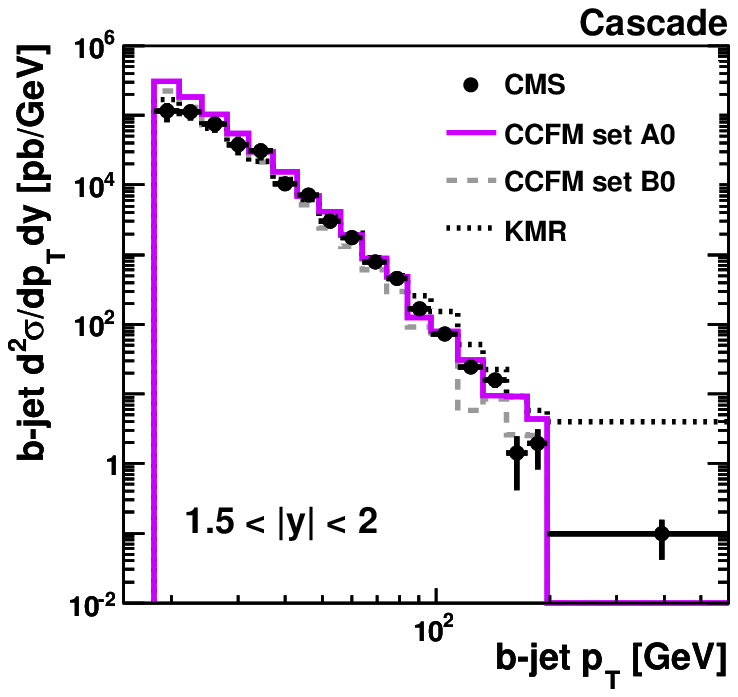, width = 8.1cm}
\caption{The double differential cross sections $d\sigma/dy\,dp_T$ of inclusive
$b$-jet production as a function of $p_T$ in different $y$ regions
calculated at $\sqrt s = 7$~TeV (\textsc{Cascade} predictions). 
Notation of all histograms is the same as in Fig.~1.
The experimental data are from CMS~\cite{2}.}
\end{center}
\label{fig6}
\end{figure}

\begin{figure}
\begin{center}
\epsfig{figure=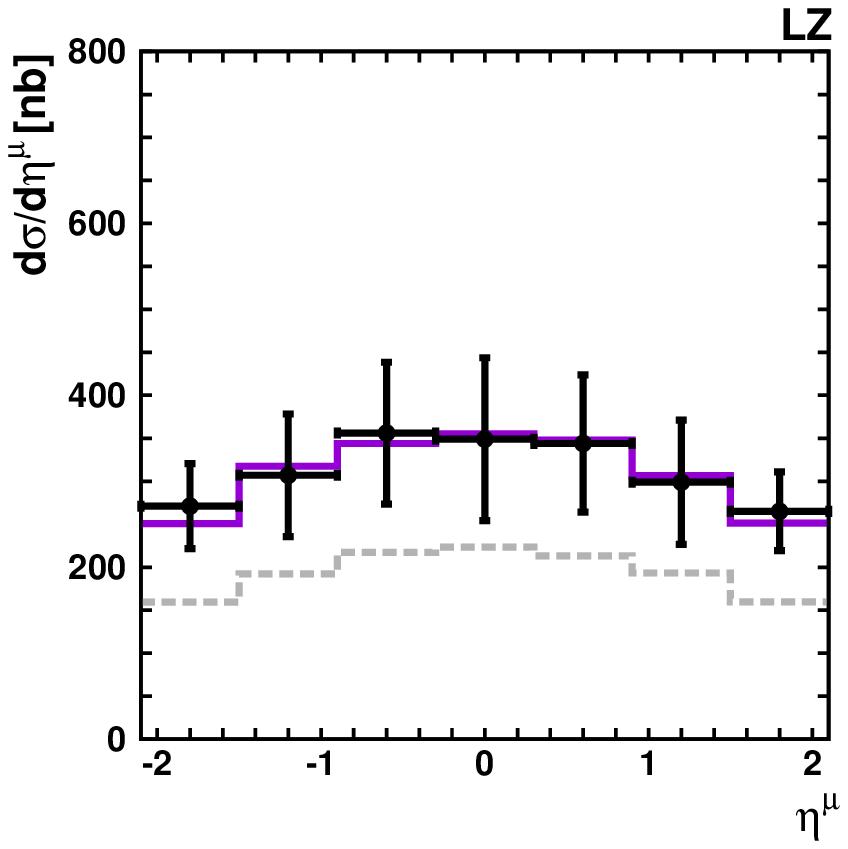, width = 8.1cm}
\epsfig{figure=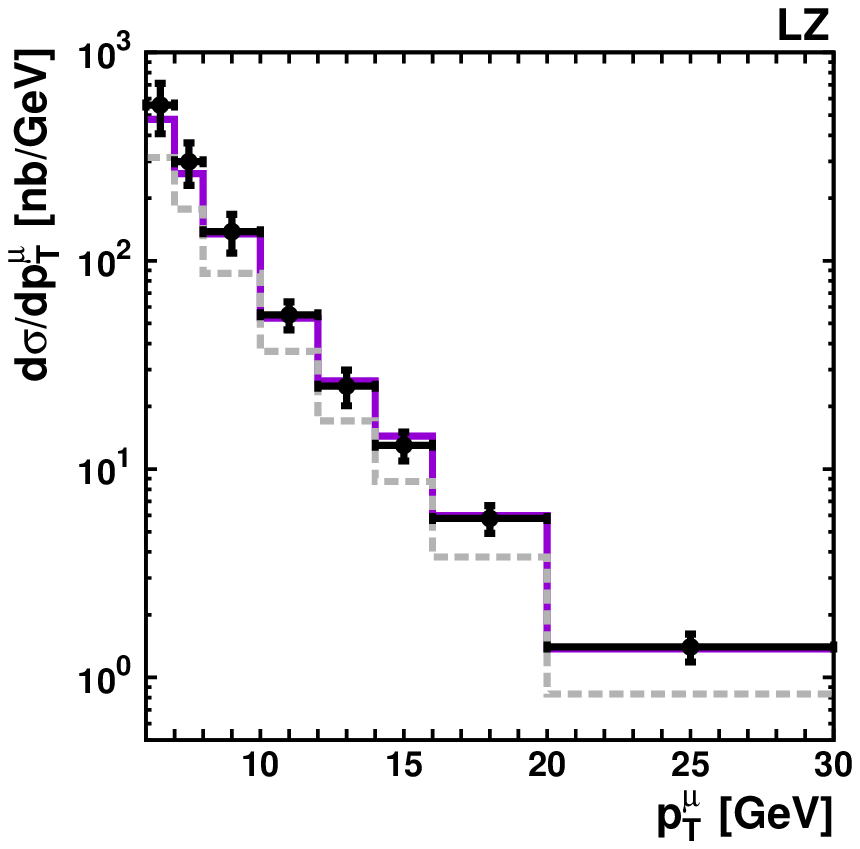, width = 8.1cm}
\caption{Importance of non-zero transverse momentum of incoming gluons in open $b$ quark production
at the LHC. The solid histograms correspond to the results 
obtained according to the master formula~(1). The dotted histograms are obtained by using 
the same formula but now we switch off the virtualities of both incoming gluons in 
partonic amplitude and apply an additional requirement ${\mathbf k}_{1,2 \,T}^2 < \mu_R^2$.
We have used here the CCFM A0 gluon for illustration.
The experimental data are from CMS~\cite{1}.}
\end{center}
\label{fig7}
\end{figure}

\begin{figure}
\begin{center}
\epsfig{figure=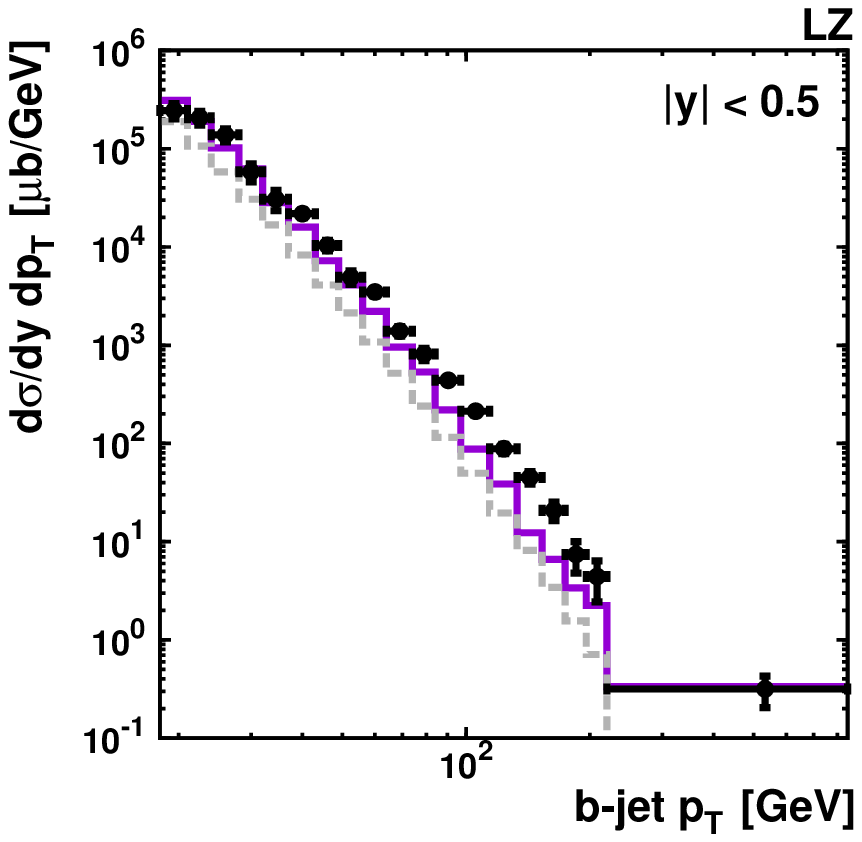, width = 8.1cm}
\epsfig{figure=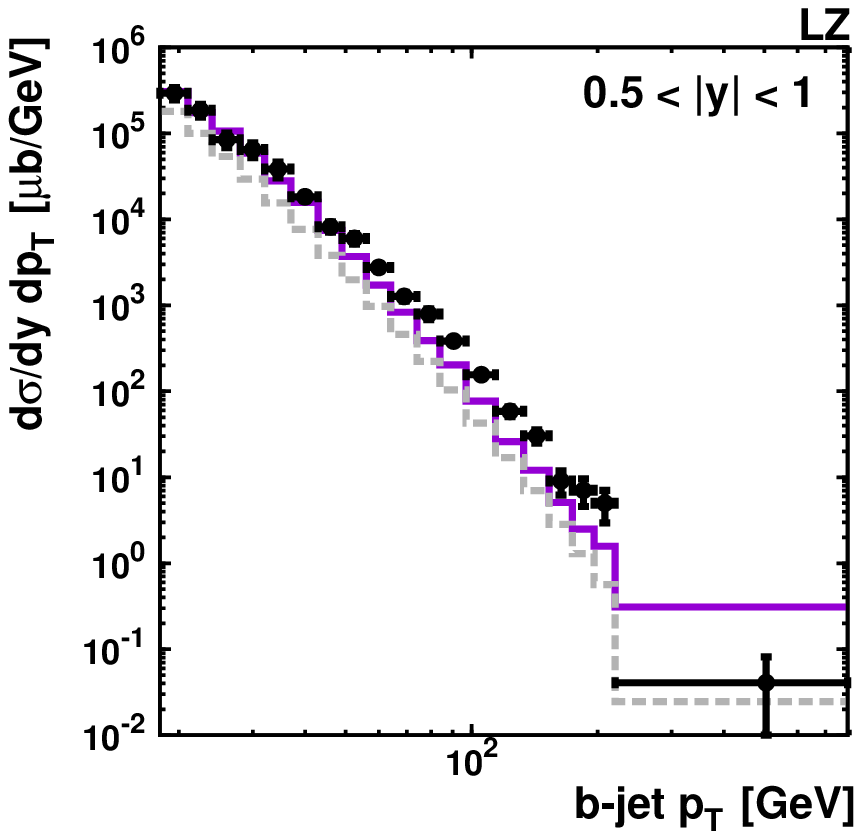, width = 8.1cm}
\epsfig{figure=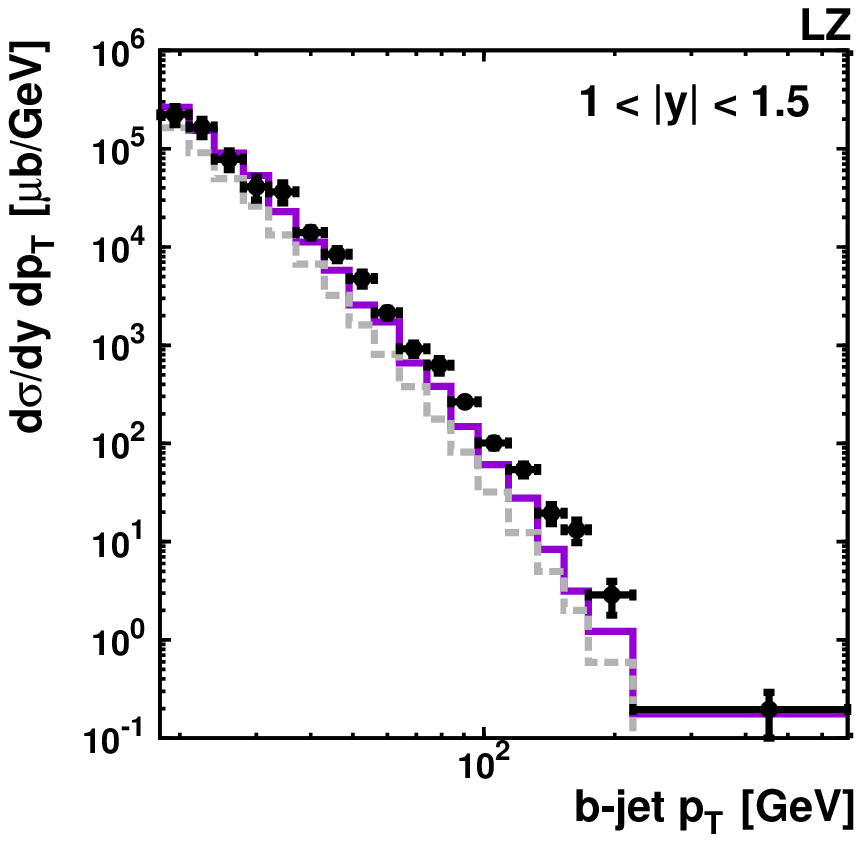, width = 8.1cm}
\epsfig{figure=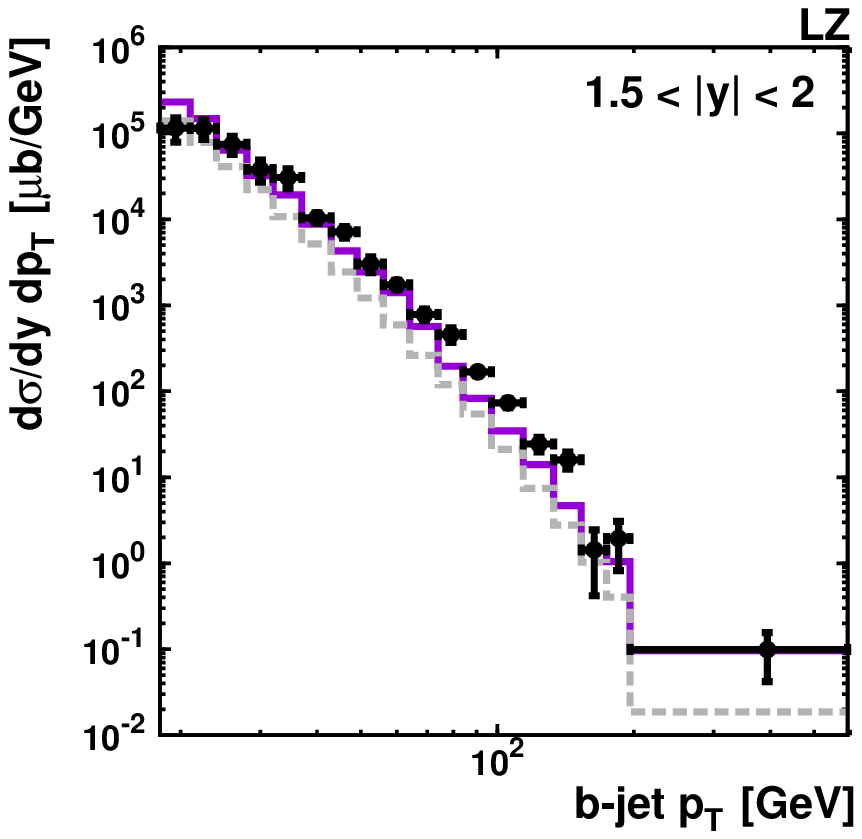, width = 8.1cm}
\caption{Importance of non-zero transverse momentum of incoming gluons in $b$-jet production
at the LHC. Notation of all histograms is the same as in Fig.~7. 
The experimental data are from CMS~\cite{2}.}
\end{center}
\label{fig8}
\end{figure}

\end{document}